\documentclass[doublecol]{epl2} 

\usepackage{amssymb}
\usepackage{amsfonts}
\usepackage{amsmath}
\usepackage{amsthm}
\usepackage{bbm}
\usepackage{xcolor}

\title{Effect of disorder and noise in shaping the dynamics of power grids}
\shorttitle{Effect of disorder and noise in power grids} 

\author{L. Tumash\inst{1} \and S. Olmi\inst{2,}\inst{3} \and E. Sch{\"o}ll\inst{1}}
\shortauthor{L. Tumash \etal}

\institute{                    
  \inst{1}Institut f{\"u}r Theoretische Physik, Technische Universit\"at Berlin, Hardenbergstra\ss{}e 36, 10623 Berlin, Germany\\
  \inst{2}INRIA Sophia Antipolis M\'editerran\'ee, 2004 Route des Lucioles, 06902 Valbonne, France\\
  \inst{3}CNR - Consiglio Nazionale delle Ricerche - Istituto dei Sistemi Complessi, 50019, Sesto Fiorentino, Italy\\
}


\abstract{The aim of this paper is to investigate complex dynamic networks which can model high-voltage power grids with renewable, fluctuating energy sources. For this purpose we use the Kuramoto model with inertia to model the network of power plants and consumers. In particular, we analyse the synchronization transition of networks of $N$ phase oscillators with inertia (rotators) whose natural frequencies are bimodally distributed, corresponding to the distribution of generator and consumer power. First, we start from globally coupled networks whose links are successively diluted, resulting in a random Erd\"{o}s-Renyi network. We focus on the changes in the hysteretic loop while varying inertial mass and dilution. Second, we implement Gaussian white noise describing the randomly fluctuating input power, and investigate its role in shaping the dynamics. Finally, we briefly discuss power grid networks under the impact of both topological disorder and external noise sources.}

\begin{document}

\maketitle

\section{Introduction}

One of the fundamental technological infrastructure requirements is a reliable supply of electric power, which affects all aspects of our life \cite{MAR08,TUR99}. 
Nowadays we are witnessing a time of global changes of power generation worldwide, which are mainly driven by concerns about climate. In particular, the reduction of power generation 
by coal-fired power plants is one of the main concepts to reduce the output of carbon dioxide, which is the main reason for global warming. Subsequently, we expect an increase of the 
inclusion of renewable energy sources into the power grid, which will lead to the reduction of greenhouse gases. An example of the transition towards sustainable energy can be seen in Germany. 
According to the so-called ``Energiewende'' one of the essential goals is to increase the fraction of renewable energy sources in the total power production to $80 \%$ in 2050. 
This implies that the power system will undergo a shift from centralized conventional to distributed power production. Thus, the new challenge will be to control lots of small generation units 
produced by power plants based on renewable energy sources instead of centrally controlling and  distributing large amounts of power from few power plants to the consumers \cite{ACK01}. 
It can lead to a strong spatial separation between power sources and consumers, since the possible locations for power plants directly depend on geographical factors. Hence the transmission 
lines should be able to carry strong loads over long distances. Another important change caused by this regime shift is the strongly fluctuating power output, since it depends on uncontrollable 
external natural factors like weather conditions \cite{MIL13,HEI10,HEI11,ANV17,ANV16}. 
Thus, one should name three major changes caused by the regime shift in the power generation. These are decentralization, spatial separation, and temporal fluctuations of the power output. 
In the present work we aim to study the stability of power grids which are affected by spatial inhomogeneity and fluctuations of the power output. 
In particular, we will investigate complex dynamic networks, which can model \textit{power grids based on renewable energy sources}. We will study not only the noise due to 
the temporal fluctuations of power, but also the effect of randomness in the network connectivity. We will discuss the similarities and differences between the temporal noise 
and topological disorder on shaping the dynamics of power grids.

The modification of the Kuramoto model by an additional inertial term was firstly reported in Refs. \cite{TAN97, TAN97a} by Tanaka, Lichtenberg, and Oishi, who were inspired
by a phase oscillator model developed by Ermentrout to mimic the synchronization mechanism observed in the firefly \textit{Pteroptix malaccae} \cite{ERM91}. Recently the model has been used 
to investigate the self-synchronization in disordered arrays of underdamped Josephson junctions \cite{TRE05}
as well as to show the emergence of explosive synchronization \cite{JI13} in a network of oscillators whose natural frequency is proportional to the node degree.
Nowadays the Kuramoto model with inertia is a standard mathematical model used to study the dynamical behavior of power plants and consumers \cite{FIL08a,FRA12,ROH12,ROH14,OLM14a,NIS15,OLM16,ROH17,GAM17}.
Most of the previous studies on power grids were devoted to networks with unimodal frequency distribution or $\delta$-distributed bimodal distributions \cite{FIL08a,ROH12,OLM14a,ROH14},
or, if bimodal distributions were considered, the network was globally coupled \cite{ACE00,OLM16}.
In the present work we will consider systems with bimodal Gaussian distribution of natural frequencies, which model energy suppliers and consumers in a more realistic way with respect 
to the previously considered cases. From the topological viewpoint we will study both globally coupled and diluted systems. A diluted system represents a network in which each node is 
connected only to some of the other nodes instead of all. This introduces some randomness into connections thereby leading to topological disorder. This has already been studied by 
Olmi et al. \cite{OLM14a} with a unimodal frequency distribution. We will demonstrate here, for the bimodal Gaussian distribution, how the results obtained for randomly diluted networks differ 
from those obtained in globally coupled networks. This enables also a comparison with the effect upon the synchronization transition caused by external white noise. 

This paper is organized as follows. First, we will define the Kuramoto model with inertia. Furthermore, we will discuss the different regimes 
occurring during adiabatic increase and decrease of the coupling strength between the nodes for a globally coupled network. Afterwards, we will incorporate random dilution and investigate changes in the synchronization transition. Then the same analysis will be performed for stochastic systems with external white noise sources. We will conclude by a comparison of diluted and noisy networks.

\section{Model}

Filatrella et al. provided the physical motivation for using the Kuramoto paradigm to model power grids \cite{FIL08a}. They distinguished two kinds of oscillators: the \textit{sources} which deliver electrical power, and the \textit{consumers} which consume this power. This means that each element of the power grid network either generates ($P^i_{source} > 0$) or consumes ($P^i_{cons} < 0$) power. Due to this, the electrical power distribution of all oscillators should be bimodal, with a maximum at $P^i_{source} > 0$ and 
one at $P^i_{cons} < 0$. In the dimensionless Kuramoto model this corresponds to a bimodal frequency distribution $\Omega_i$. Hence, throughout this work we will use bimodal distributions of frequencies constructed by superposition of two Gaussians with peaks at positive and negative frequencies, respectively. The Kuramoto model with inertia reads
\begin{equation}\label{EQ:1}
m \ddot\theta_i + \dot\theta_i = \Omega_i + \frac{K}{N_i} \sum\limits_{j=1}^N A_{ij} \sin\left(\theta_j - \theta_i\right),
\end{equation}
where $\theta_i$ and $\dot\theta_i = \omega_i $ are the instantaneous phase and frequency deviation of the $i$-th oscillator, $i=1,...,N$, relative to the collective grid frequency. 
The mass $m > 0$ indicates the value of inertia of generators and loads. $K$ is the coupling constant of the network, it describes the strength of the connectivity between the nodes. 
$A$ is the connectivity matrix, whose entries $A_{ij}$ can be either one or zero if the link between the oscillators $i$ and $j$ is present or absent, respectively. 
It is a symmetric matrix $A_{ij} = A_{ji}$, since our network represents an undirected graph. $N_i$ is the node degree of the $i$-th element, which denotes the number of the links emanating from this node. 
In case of globally coupled networks, $A_{ij} = 1$ and $N_i = N-1 \approx N$, while for randomly diluted networks the degree is set to be constant for each oscillator $N_i = N_c$. 
We introduce a dilution parameter $p=\frac{N_{c}}{N}$, which denotes the ratio of actual links per node to the number of all possible links.
Finally $\Omega_i$ represents the natural frequency of the oscillator $i$, and its value is chosen according to the probability density
\begin{equation}\label{EQ:2}
g(\Omega) = \frac{1}{2\sqrt{2 \pi}} \left[ e^{-\frac{(\Omega - \Omega_0)^2}{2}} + e^{-\frac{(\Omega + \Omega_0)^2}{2}} \right],
\end{equation}
where $g(\Omega)$ is the superposition of two Gaussians with unit standard deviation and mean values $\pm\Omega_0$. 
Thus, their peaks are located at a distance $2\Omega_0$.  In the following we will choose  $\Omega_0=2$, i.e., the Gaussians have almost no overlap.
 
In order to measure the level of synchronization between the oscillators, we introduce the complex order parameter
\begin{equation}\label{EQ:3}
r(t) e^{i\phi(t)} = \frac{1}{N} \sum\limits_{j=1}^N e^{i\theta_j},
\end{equation}
where its modulus $ r(t) \in \left [ 0 , 1 \right ] $ and argument $ \phi(t) $ indicate the degree of synchrony and mean phase angle, respectively. 
In the following we will denote $ r(t) $ as \textit{global order parameter}. An asynchronous state is characterized by $r \approx 0$, while $r=1$ corresponds to the fully synchronized solution. 
Intermediate values of $r$ correspond to states with partial or cluster synchronization. 



\section{Methods}

In this paper we will perform simulations by sweeping up and down the coupling strength $K$, following two different protocols.
Namely, with \textit{protocol (I)} we denote the procedure where the system is initialized randomly for zero coupling 
(we set random initial conditions both for phases $\left \{ \theta_i \right \}$ and frequencies $\left \{ \omega_i \right \}$ for each node $i$). Then the coupling is increased in steps of $\Delta K = 0.5$, until the maximum coupling  $K_M$ is reached. Thereby, the global order parameter of the system increases, and the maximum coupling corresponds to the achievement of synchronization. In the case of nonzero coupling we initialize the system by taking the last configuration from each previous simulation for all the following steps $\Delta K$. At each step the simulations are running for a transient time $T_R$ followed by a time $T_W$, over which the average values of the global order parameter $\bar{r}$, 
the velocities $\left \{ \bar{\omega}_i\right \}$ and the maximum natural frequency of the locked oscillators $\Omega_M$ are calculated. 
\textit{Protocol (II)} denotes the reverse procedure. The initial state corresponds to a frequency-synchronized system. The simulations are also performed adiabatically, i.e., the system is always initialized according to the last configuration. We decrease the coupling parameter in steps of $\Delta K = 0.5$, until the system becomes uncoupled and completely asynchronous. 

\begin{figure}[]
\begin{minipage}[h]{0.93\linewidth}
\center{\includegraphics[width=\linewidth]{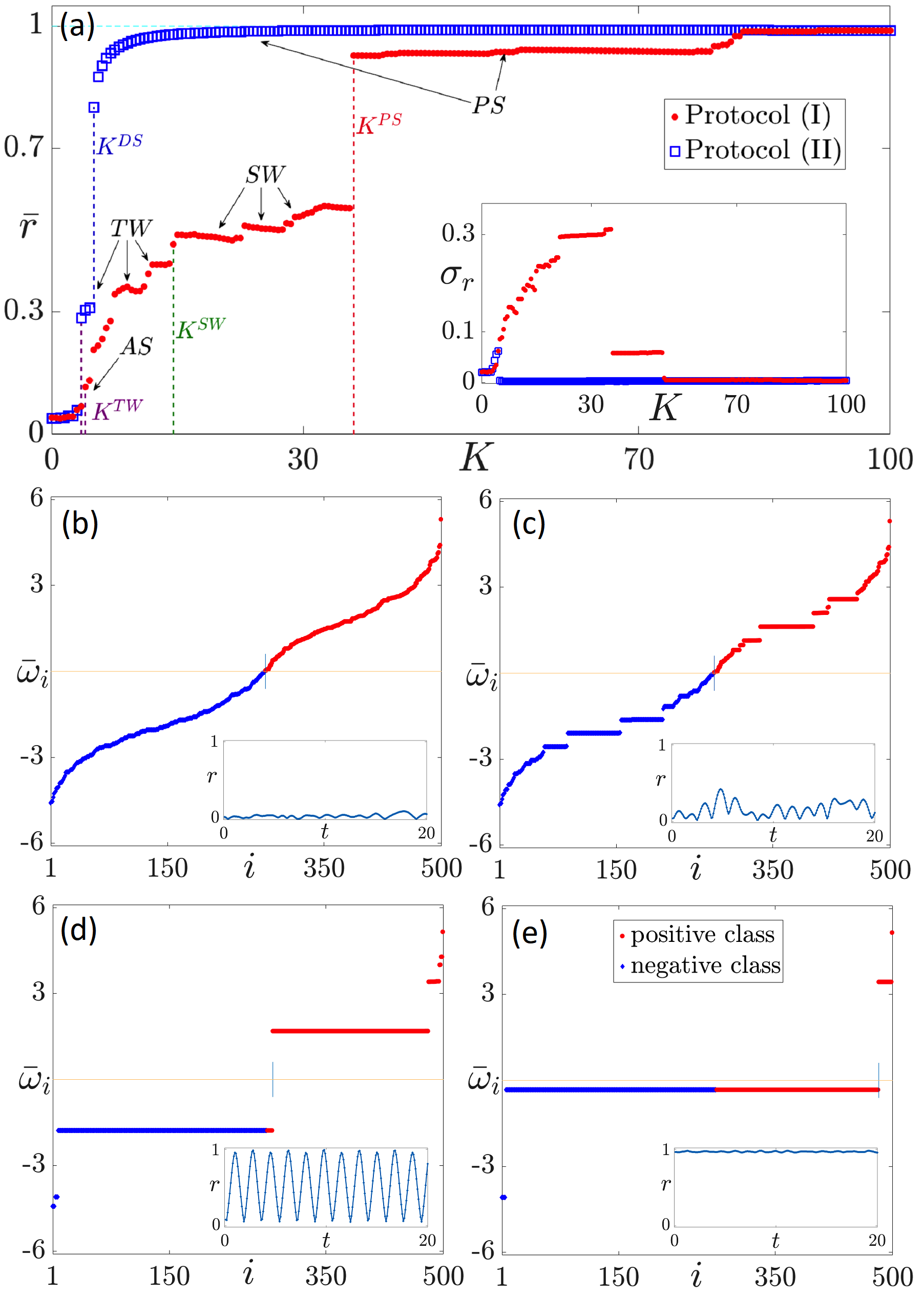}}
\end{minipage}
\caption{(a): Time-averaged global order parameter $\bar{r}$ as a function of coupling constant $K$ for two series of simulations, obtained by following the protocol (I) 
(upsweep, red filled circles) and (II) (downsweep, blue empty squares) for global coupling. The vertical dotted lines denote the critical values of coupling $K$ for traveling waves ($K^{TW}$, blue), 
standing waves ($K^{SW}$, purple), partial synchronization ($K^{PS}$, green) and the value at which desynchronization occurs ($K^{DS}$, yellow). 
Inset: standard deviation of the global order parameter $\sigma_{r}$ vs $K$. Average phase velocity $\bar{\omega}_i$ as a function of node $i$ 
for (b) $K = 2$, $\bar{r} = 0.046$ (asynchronous state); (c) $K = 5$, $\bar{r}=0.167$ (traveling wave); (d) $K = 34$, $\bar{r} = 0.556$ (standing wave); (e) $K=60$, $\bar{r}=0.941$ (partial synchronization). 
The nodes are labeled such that the average phase velocities $\bar{\omega}_i$ are sorted from low to high values. The insets in panels (b)-(e) illustrate $r(t)$ within a time interval $t \in (0, \, 20)$.
Parameters: $m=8$, $p=1.0$, $N=500$, $T_R = 5000$, $T_W = 200$.}
\label{FIG:1}
\end{figure}

In Fig.~\ref{FIG:1}a the average global order parameter $\bar{r}$ is depicted as a function of coupling $K$, obtained by following protocols (I) and (II). 
In the inset the standard deviation $\sigma_r$ of $r(t)$ is shown as a function of $K$.
Following the sequence of simulations corresponding to protocol (I) we see that the system first exhibits an asynchronous state (AS), where the 
average order parameter $\bar{r} \approx 1/\sqrt{N}$ while its time-dependent behavior is irregular and the oscillators are characterized by different average phase velocities $\bar{\omega}_i$ 
(see Fig.~\ref{FIG:1}b).  
The system remains asynchronous up to a critical value $K^{TW}$, when the average global order parameter $\bar{r}$ exhibits a jump to higher values such that $\bar{r} > 0.1$. 
At this critical coupling, chaotically whirling oscillators suddenly get locked into one or more clusters of 
nodes drifting together with the same average phase velocity $\bar{\omega}_i$ (see Fig.\ref{FIG:1}c). These clusters are defined by their \textit{maximum locking frequency $\Omega_M$} and 
the number of locked oscillators $N_L$. The coexistence of locked nodes with chaotically whirling oscillators corresponds to the traveling wave (TW) solution. 
Here $r(t)$ exhibits irregular temporal oscillations (see inset). According to Olmi et al \cite{OLM14a}, the amplitude of $r(t)$ depends approximately linearly on the number of oscillators in drifting clusters.
As the coupling parameter $K$ continues to increase, the clusters are growing by adding more chaotically drifting oscillators. 
Thereby the oscillation amplitude of the corresponding global order parameter $r(t)$ increases. The clusters continuously grow until the whole network consists 
of only two symmetric clusters of locked oscillators drifting together with opposite average phase velocities equal to $\bar{\omega}_i \approx \pm \Omega_0$ (see Fig.\ref{FIG:1}d). 
In this case the number of unlocked oscillators $N-N_L$ is vanishingly small comparing to the total network size $N$. The corresponding value of coupling required to pass into such a 
state is denoted in Fig.~\ref{FIG:1}a as $K^{SW}$. This regime is known as a standing wave (SW) solution. The oscillations of $r(t)$ achieve their maximum amplitude and become almost periodic (see inset), 
while their period is related to the average frequency of the clustered oscillators $|\Omega_0|$ \cite{OLM16}.
In this case the average global order parameter is typically equal to $\bar{r} \approx 0.5$. The network behaves as two distinct subnetworks each corresponding to a unimodal Gaussian 
distribution of frequencies with opposite peaks: one is located at $-\Omega_0$ (loads) while the other one is centered at $+\Omega_0$ (generators). 
The coupling parameter $K$ is further increased in steps $\Delta K$ until the critical value $K^{PS}$ is reached. At this value two drifting symmetric clusters become unstable and merge 
into a unique stationary cluster with $\bar{\omega}_i \approx 0$ (see Fig.\ref{FIG:1}e). This transition to partial synchronization (PS) can be easily identified within the synchronization 
transition of any network (independently of the topology and the disorder) since at the critical coupling $K^{PS}$ the average global order parameter $\bar{r}$ exhibits a rapid discontinuous jump, 
thereby almost doubling its value ($\bar{r} > 0.9$). The system smoothly approaches the fully synchronized regime by further increase of coupling $K$. 
The corresponding global order parameter $r(t)$ becomes stationary.

If we perform protocol II now, the system loses synchrony, i.e., desynchronizes, for a coupling value $K^{DS}\neq K^{PS}$.
It is remarkable that $K^{DS} < K^{PS}$, which is a clear indication of \textit{the hysteretic nature of the synchronization transition}. Let us introduce $W_h$, which will be used 
to indicate the width of the hysteretic region, i.e., $W_h =  K^{PS}-K^{DS}$. In the case depicted on Fig.~\ref{FIG:1}a we obtain $W_h = 25.5$. Relating to the bifurcation analysis of the Kuramoto 
mean-field performed by Tanaka et al. \cite{TAN97,TAN97a}, this pattern is the numerical manifestation of bistability. By using their notation we define 
$ \frac{4}{\pi} \sqrt{\frac{K \bar{r}}{m}} = \Omega_P $, $ K \bar{r} = \Omega_D$. There are two fixed points for small $K$ ($\Omega_M < \Omega_P + \Omega_0$): a stable node and a saddle. 
At larger coupling values $K^{TW}<K<K^{PS}$ the maximum $\Omega_M$ becomes also larger ($\Omega_M> \Omega_P + \Omega_0$) and a limit cycle emerges via a homoclinic bifurcation.
This corresponds to the coexistence of locked and oscillatory solutions (stable limit cycle coexisting with the stable node). By further increasing $K$ the system reaches the pure oscillatory solution for 
$\Omega_M > \Omega_D$ due to a saddle-node bifurcation, in which only the limit cycle persists. However, if we perform simulations by following protocol (II), the system remains stationary until 
the stable fixed point solution completely disappears. This happens if we decrease $K$ and achieve $\Omega_M >\Omega_D + \Omega_0$.  Thereby, the system reveals oscillatory behavior due to the limit cycle. 
Thus we see that the hysteretic loop $W_h$ depends on the inertial mass $m$ while running protocol (I). In contrast to that the inertial mass does not seem to play any role for protocol (II). 
The observed behavior can be explained by the fact that for $K < K^{PS}$ ($K < K^{DS}$) for protocol (I) (protocol (II)), the network behaves as two independent sub-networks each characterized by a 
unimodal frequency distribution, centered at $\pm \Omega_0$, respectively, while, for sufficiently large coupling strength, once the system exhibits only one large cluster with zero velocity, 
the network behaves like a single entity, as for the case with a unimodal distribution centered at zero \cite{TAN97,TAN97a}.

\section{Deterministic dynamics in diluted networks}

\begin{figure}[]
\begin{minipage}[h]{0.93\linewidth}
\center{\includegraphics[width=\linewidth]{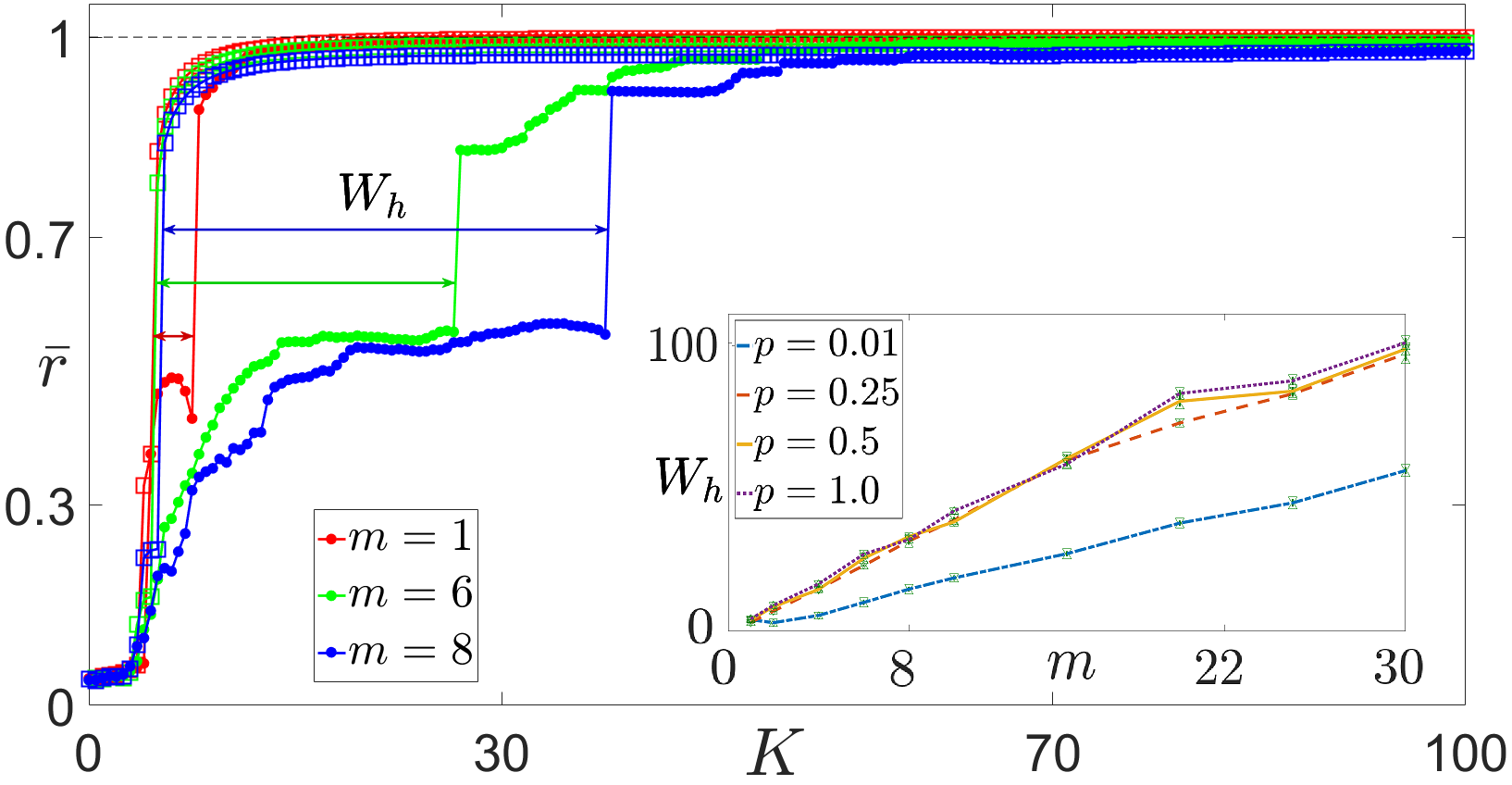}}
\end{minipage}
\caption{(a): Time-averaged global order parameter $\bar{r}$ as a function of coupling constant $K$ for two series of simulations, performed by following protocols (I) and (II), denoted by filled circles (upsweep) and empty squares (downsweep), respectively, for a network with fixed dilution parameter $p=0.25$ and mass $m=1$ (red);  $m=6$ (green); $m=8$ (blue). Colored arrows indicate the width $W_h$ of the corresponding hysteretic region. Inset: $W_h$ as a function of $m$ for $p=1.0$ (purple), $p=0.50$ (yellow), $p=0.25$ (red), $p=0.01$ (blue). Other parameters as in Fig.~\ref{FIG:1}.}
\label{FIG:2}
\end{figure}

In this section we consider synchronization transitions of spatially disordered (diluted) networks with deterministic dynamics for different values of mass $m$ and dilution $p$. 
First of all, we investigate how the width of the hysteretic loop $W_h$ changes if we fix the dilution parameter $p=0.25$ and vary the mass ($m=1$, $m=6$ and $m=8$), see Fig.~\ref{FIG:2}a. 
For all the masses we observe a clear hysteretic behaviour obtained by simulations following protocols (I) and (II). As expected, the width of the hysteretic loop $W_h$ depends on the inertial mass $m$. 
As discussed above for globally coupled networks, smaller masses increase the lower boundary of the interval $\frac{4}{\pi} \sqrt{\frac{K \bar{r}}{m}}$, which leads to contraction of this interval. 
Also here, in diluted networks, we observe a tiny hysteretic loop $W_h$ for $m=1$, while a large mass $m=8$ provides strong hysteresis with large $W_h$. This is clear since the system is governed by the 
second order differential equation $m \ddot{\theta} = f \left ( \theta, \, \dot{\theta} \right )$. The acceleration term $\ddot{\theta} = \frac{f \left ( \theta, \, \dot{\theta} \right )}{m}$ 
is inversely proportional to the inertia. Thus, increasing mass slows down the oscillations and we obtain larger values of critical coupling required to pass to partial synchronization. 
The summary of the results for different masses in the range $1 \le m \le 30$ is given in the inset of Fig.~\ref{FIG:2}a, indicating that this tendency holds even for highly diluted systems with $p=0.01$.

\begin{figure}[]
\begin{minipage}[h]{0.9\linewidth}
\center{\includegraphics[width=\linewidth]{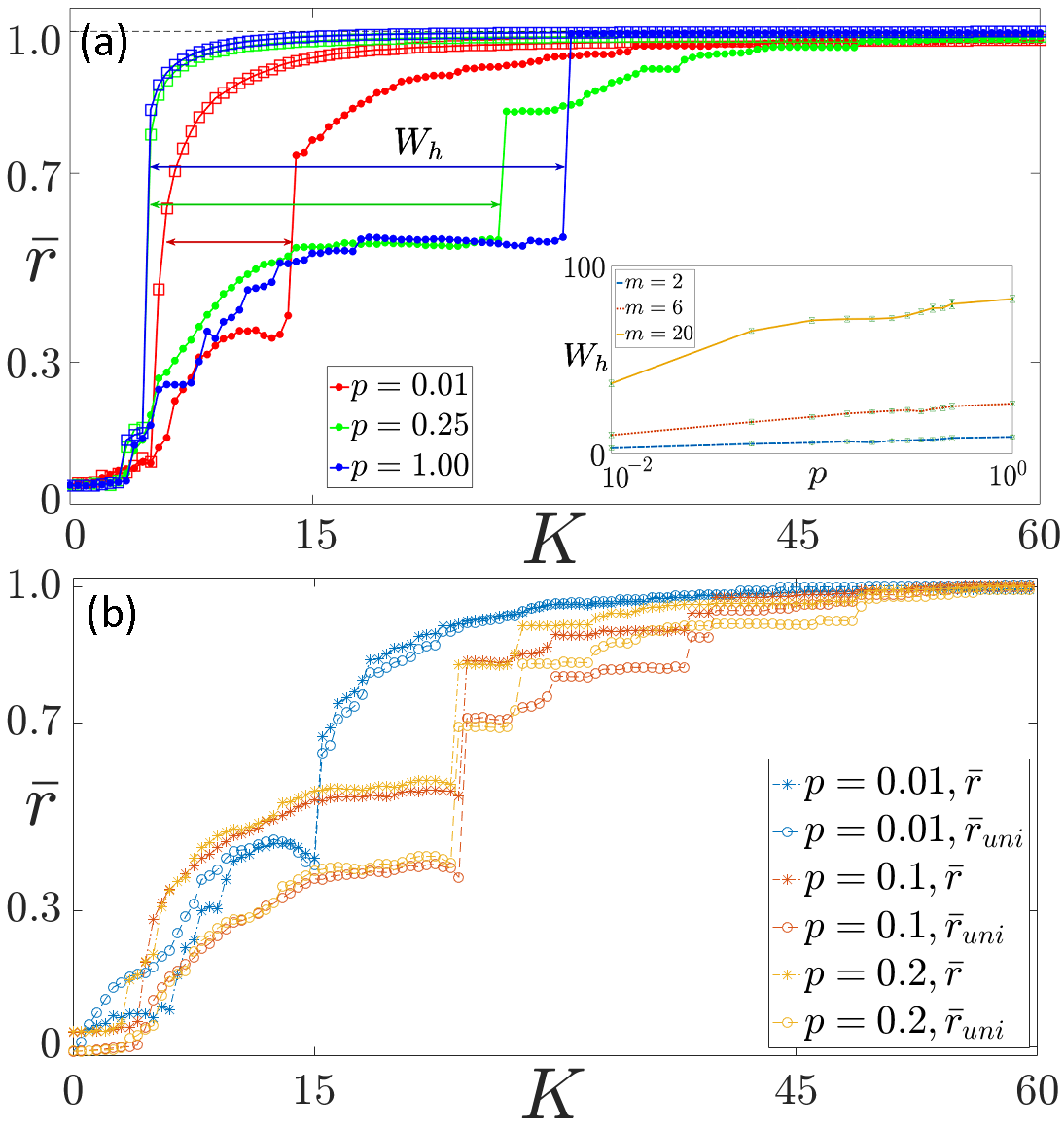}}
\end{minipage}
\caption{Time-averaged global order parameter $\bar{r}$ as a function of coupling strength $K$ for a network with fixed mass $m=6$ and dilution $p=0.01$ (red);  $p=0.25$ (green); $p=1.0$ (blue). Colored arrows indicate the width  $W_h$ of the corresponding hysteretic region. 
Inset: $W_h$ as a function of $p$ for three different masses $m=2$ (blue), $m=6$ (red), and $m=20$ (yellow). Note that the abscissa is in a logarithmic scale.  (b): Time-averaged global order parameter 
$\bar{r}$ (asterisks) and universal global order parameter $\bar{r}_{uni}$ (empty circles) vs $K$ (upsweep) for $p=0.01$ (blue), $p=0.1$ (orange) and $p=0.2$ (yellow). Other parameters as in Fig.~\ref{FIG:1}.}
\label{FIG:3}
\end{figure}

Furthermore, we investigate the changes in the synchronization transitions caused by topological disorder within the network. We fix $m=6$ and consider three different topological situations: 
global coupling ($p=1.0$), $p=0.25$ and $p=0.01$ (Fig.~\ref{FIG:3}). Also here we observe hysteretic behavior for all considered parameters.
However, we can see that \textit{the hysteretic loop decreases as the network topology becomes more sparse}. 
When we increase the dilution, i.e., decrease $p$, sustaining two symmetric clusters drifting together with opposite average phase velocites 
$ \bar{\omega}_i = \pm 2$ (standing waves) becomes quite hard. Thus, the hysteretic region decreases. We can also see that high disorder affects the width $W_h$ by slightly increasing the critical coupling $K^{DS}$ at which the system collapses towards asynchrony by following protocol II. Olmi et al. have already shown for a unimodal frequency distribution \cite{OLM14a} that the synchronization transition remains hysteretic even for highly disordered networks up to a very high level of dilution, which corresponds to few links per node only. In the inset of Fig.~\ref{FIG:3} we see that the hysteretic region increases as we increase the average 
connectivity of the network, for all investigated values of inertia.

\begin{figure}[]
\begin{minipage}[h]{1.0\linewidth}
\center{\includegraphics[width=\linewidth]{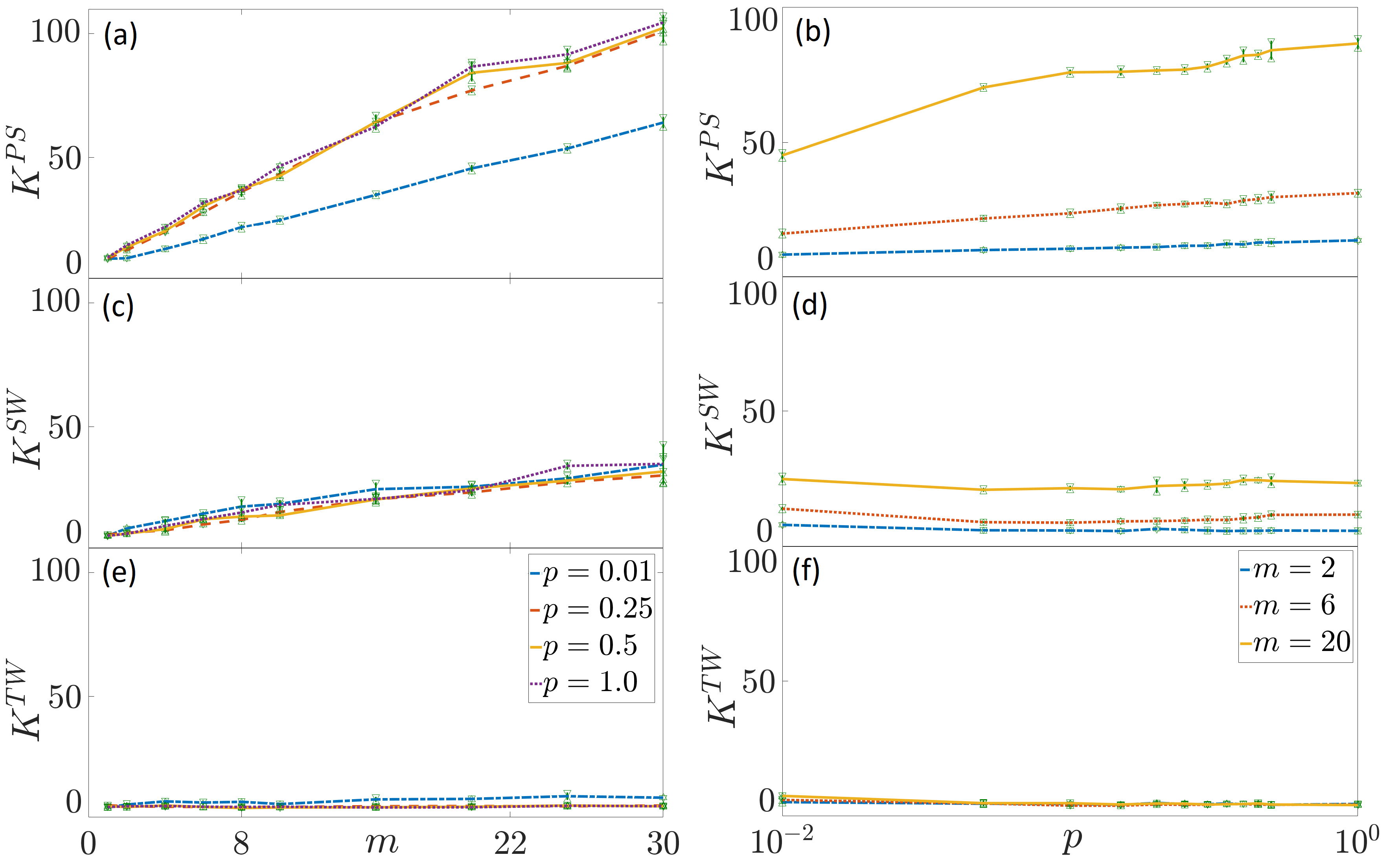}}
\end{minipage}
\caption{Critical coupling for different regimes as a function of mass (left column) and dilution (right column). The curves in the left column are presented for four different network topologies: $p=1.0$ (purple), $p=0.50$ (yellow), $p=0.25$ (red), $p=0.01$ (blue), and in the right column for three different masses: $m=2$ (blue), $m=6$ (red), $m=20$ (yellow).  The critical values of coupling are depicted for the following regimes: 
(a,b) Partial frequency synchronization ($K^{PS}$); (c,d) Standing wave ($K^{SW}$); (e,f) Traveling wave ($K^{TW}$). Other parameters as in Fig.~\ref{FIG:1}.}
\label{FIG:4}
\end{figure}

The main features of the synchronization transition profile are preserved also if we perform the upsweep analysis with respect to the universal order parameter $\bar{r}_{uni}$
introduced by Schr\"{o}der et al. \cite{SCH17a} as a generalization of Eq.~\eqref{EQ:3}, including the influence of the network topology. In particular, $\bar{r}_{uni}$ and $\bar{r}$ are both depicted in Fig. 3b as functions of $K$. Comparing these two curves, $\bar{r}_{uni}$ generally shows lower values than $\bar{r}$ at the same $K$ (especially for not too low dilution parameter). 
Also $\bar{r}_{uni}$ is non-monotonic with increasing $K$, and shows discontinuous jumps and plateaus corresponding to intermediate states.

To sum up the results for deterministic dynamics, Fig.~\ref{FIG:4} depicts the critical coupling strength required to pass from a particular state to another as a function of mass $m$ (left column) and dilution $p$ (right column). While the critical coupling values at which traveling waves ($K^{TW}$) or standing waves ($K^{SW}$) emerge essentially do not depend on the dilution level, the critical coupling at which the system reaches partial synchronization $K^{PS}$ depends on the average connectivity of the network, and a smaller coupling strength is required for very low connectivity. Moreover, this dependence is enhanced for larger values of inertia.
On the other hand, if we keep the dilution fixed, the critical coupling increases with increasing mass; this tendency is particular evident for $K^{PS}$ and it is enhanced for increasing values of connectivity.

\section{Stochastic dynamics with fluctuating power}
In this section we investigate the influence of power fluctuations of generators, in particular due to renewable energy, and consumers. 
In a Langevin approach we add external noise sources:
\begin{equation}\label{EQ:7}
m \ddot\theta_i + \dot\theta_i = \Omega_i + \frac{K}{N_i} \sum\limits_{j=1}^N A_{ij} \sin\left(\theta_j - \theta_i\right) + \sqrt{ 2D} \xi_i(t),
\end{equation}
where $ \xi_i (t)$ is independent Gaussian white noise with  $ \langle \xi_i \rangle = 0$ and $ \langle \xi_i (t) \xi_j (s) \rangle = \delta_{ij}\delta(t-s)$, and $D$ is the noise intensity. 

\begin{figure}[]
\begin{minipage}[h]{0.93\linewidth}
\center{\includegraphics[width=\linewidth]{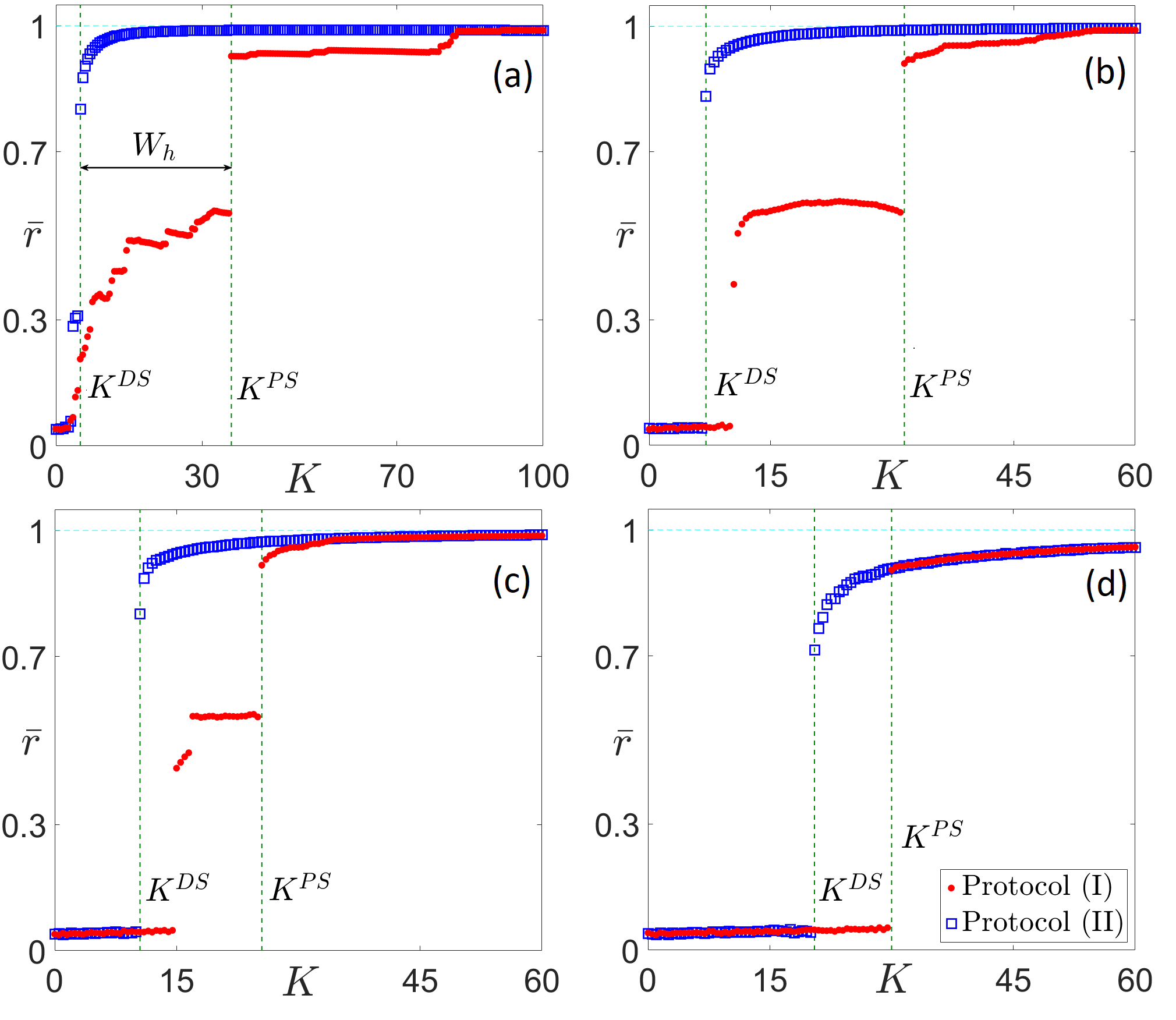}}
\end{minipage}
\caption{Time-averaged global order parameter $\bar{r}$ versus coupling constant $K$ for a globally coupled network with stochastic dynamics, following protocol 
(I) (red filled circles) and (II) (blue empty squares) for $m=8$ and noise intensities (a) $\sqrt{2D}  = 0$;  (b) $\sqrt{2D}  = 9$; (c) $\sqrt{2D}  = 15$; 
(d) $\sqrt{2D}  = 30$. The vertical green dashed lines indicate the boundaries of the hysteretic region. Other parameters as in Fig.~\ref{FIG:1}.}
\label{FIG:5}
\end{figure}

The synchronization transition scenarios for stochastic dynamics with different noise intensities $\sqrt{2D}$ and fixed mass $m$ are depicted as a function of coupling strength $K$ in Fig.~\ref{FIG:5}. In order to clearly separate the effects of topological disorder and temporal noise we set $p=1$ (global coupling). We can observe that the hysteresis (bounded between the vertical dashed green lines) notably decreases with increasing noise intensity $\sqrt{2D}$. 
It is also remarkable that large noise prevents intermediate states (traveling and standing wave). 
In particular, in panel (d) we observe a synchronization transition from the asynchronous state directly to partial synchronization, without passing through intermediate cluster states. Thus, we can conclude that large stochastic fluctuations perturb the whole network, preventing it for small coupling from leaving the asynchronous state. However, if the coupling becomes sufficiently strong, the system jumps into partial synchronization (almost full synchronization).
This can also explain why the system requires a smaller critical coupling strength to pass into partial synchronization $K^{PS}$ compared to the cases in which intermediate cluster states appear. This is an intriguing example of the \textit{constructive role of noise}. Note also that intermediate values of noise intensity have a comparable effect upon the dynamics as intermediate values of dilution.

\begin{figure}[]
\begin{minipage}[h]{0.93\linewidth}
\center{\includegraphics[width=\linewidth]{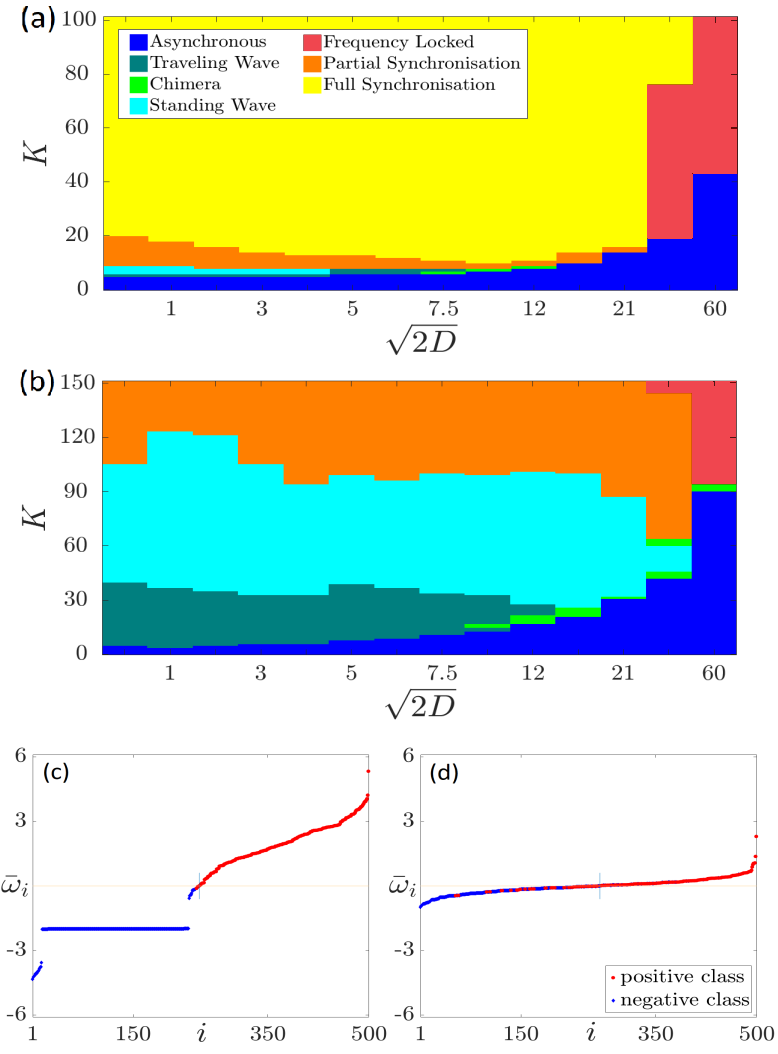}}
\end{minipage}
\caption{Map of regimes in the $ \left ( \sqrt{2D}, \, K \right ) $ plane for (a) $m=1$; (b) $m=30$ for a globally coupled network with stochastic dynamics: asynchronous state (dark blue), traveling wave (dark green), noise-induced chimera (light green), standing wave (cyan), frequency locked (red), partial frequency synchronization (orange), full synchronization (yellow). Average phase velocity $\bar{\omega}_i$ as a function of node $i$ of (c) Chimera state ($\bar{r}=0.393$: $\bar{r}_{-} = 0.818$ and $\bar{r}_{+} = 0.0638$; $m=8$; $\sqrt{2D}=9$, $K=11$); (d) Frequency-locked state ($\bar{r}=0.618$: $\bar{r}_{-} = 0.620$ and $\bar{r}_{+} = 0.624$; $m=1$; $\sqrt{2D} = 60$, $K=48$). Other parameters as in Fig.~\ref{FIG:1}.}
\label{FIG:6}
\end{figure}

Furthermore, we present maps of regimes in the $\left ( \sqrt{2D}, \, K \right ) $ plane for two different masses $m=1$ and $m=30$ (Fig.~\ref{FIG:6}(a) and (b); upsweep). This provides general insight into the interplay of noise intensity $\sqrt{2D}$, coupling $K$, and mass $m$ in shaping the dynamics of the system. We observe that the dark blue area (asynchronous regime) becomes larger as the noise intensity $\sqrt{2D}$ increases. A network with small $m$ is usually characterized by a small hysteretic region, in which we mostly even do not observe traveling waves. Moreover, we observe a tiny light green island corresponding to \textit{chimera} states \cite{CHIM03,CHIM04,CHIM07,CHIM13,CHIM14,CHIM08,CHIM09,CHIM02} for $7.5 <\sqrt{2D}< 12$ 
(Fig.~\ref{FIG:6}c). This state is defined by the spatial coexistence of a frequency synchronized part (characterized by a partial order parameter $\bar{r}_{-}$) and a completely incoherent part (characterized by a partial order parameter $\bar{r}_{+}$). In our case the two parts correspond to loads (negative $\Omega$ class) 
and generators (positive $\Omega$ class), respectively. There is also a noise-induced \textit{frequency-locked} regime for large noise intensities (Fig.~\ref{FIG:6}d). This regime is characterized by high values of $\bar{r}$ although the corresponding $\bar{\omega}_i$ profile is not constant but slightly tilted. Finally, an extended parameter region is characterized by the partial and full synchronization states that are easily accessible for low values of inertia.
On the other hand, for $m=30$ the influence of white noise is weaker; it is not capable to destroy the intermediate cluster states. Therefore the map of regimes exhibits a very large region where traveling and standing waves are observable. Noise-induced chimera and frequency-locked states are observable also in this limit of high value of inertia. Synchronization is more difficult to reach due to the large $m$ value.

Finally, we have also examined a network under the simultaneous impact of both topological disorder and external white noise. 
If an intermediate level of noise is chosen, we find that, irrespectively of the used dilution $p$, the synchronization transition profile remains hysteretic. However, the synchronization becomes more difficult to achieve as we increase noise. 
\section{Conclusions}

We have investigated how topological disorder and temporal power fluctuations affect the scenarios of synchronization transitions in power grids. Dilution of the links (topological disorder), 
noise (temporal fluctuations), and small values of inertia are typical for power grids with renewable energy sources as opposed to conventional generators.
We have verified the hysteretic behaviour of the synchronization transition for different inertial masses and dilution of the network connectivity. 
The system reveals larger hysteretic loops as we either increase inertial mass or reduce topological disorder. The cluster states occurring during the transition from complete incoherence to 
complete coherence with increasing coupling strength were classified and illustrated: traveling waves, standing waves and partial synchronization. 
Finally, we have presented critical values of the coupling strength required to pass into those various states in dependence of dilution and mass. The most evident tendency was observed 
for partial synchronization, which requires stronger coupling for larger values of mass and weaker dilution. Since synchronized states are mandatory for stable operation of power grids, 
our work suggests that special attention must be paid in tuning the capacity of an existing grid or designing new power grid networks. 

Moreover, we have considered the Kuramoto model with white noise modelling stochastic power fluctuations typical for renewable energies \cite{SCH17, GAM17}, and computed the synchronization 
transition scenarios. Similar to topological disorder (dilution), noise leads to a decrease of the hysteretic region within the synchronization scenario. In contrast to dilution, larger noise intensities 
are able to induce incoherence in the velocities and their average values within one cluster, giving rise to a \textit{frequency-locked} state. For intermediate noise intensities, noise can induce 
\textit{chimera} states. In particular we have provided maps of regimes of two systems with different values of inertia under the impact of noise. We have found that noise suppresses intermediate cluster states, 
and, for large noise intensities, the synchronization scenario proceeds directly from the asynchronous state to the almost fully synchronized state. Finally, we have found that 
\textit{noise may have a constructive role} in decreasing the critical coupling strength necessary to reach partial or full synchronization.

In the future it would be challenging to extend this analysis to more realistic types of noise, e.g., intermittent or non-Gaussian, and more realistic natural frequency distributions. 
For long distances between power generators and consumers delay effects should be taken into account.

\acknowledgments
This work was supported by DFG in the framework of SFB 910.


\end{document}